\documentclass{JHEP}
\usepackage{epsfig}

\usepackage{amssymb}

\def\drawbox#1#2{\hrule height#2pt
        \hbox{\vrule width#2pt height#1pt \kern#1pt
              \vrule width#2pt}
              \hrule height#2pt}

\def\Fund#1#2{\vcenter{\vbox{\drawbox{#1}{#2}}}}
\def\Asym#1#2{\vcenter{\vbox{\drawbox{#1}{#2}
              \kern-#2pt       
              \drawbox{#1}{#2}}}}

\def\funda{\Fund{6.5}{0.4}}
\def\asymm{\Asym{6.5}{0.4}}

\def\symm{\funda\kern-0.4pt\funda}

\def\tr{\mathop{\rm tr}}

\newcommand{\half}{\frac{1}{2}}

\newcommand{\R}{\ensuremath{\mathbb R}}

\newcommand{\Z}{\ensuremath{\mathbb Z}}

\newcommand{\RP}[1]{\ensuremath{{\R}{\rm P}^{#1}}}
\newcommand{\SSS}{\ensuremath{\mathbb S}}
\newcommand{\I}{\ensuremath{\mathbb I}}
\newcommand{\J}{\ensuremath{\mathbb J}}
\newcommand{\QQ}{\ensuremath{\mathbb Q}}
\newcommand{\TT}{\ensuremath{\mathbb T}}

\newcommand{\hatplane}[2]{\widehat{\mbox{O}{#1}}^
           {\raisebox{-4.5pt}{$\scriptstyle {#2}$}}}
\newcommand{\tildeplane}[2]{\widetilde{\mbox{O}{#1}}^
           {\raisebox{-4.5pt}{$\scriptstyle {#2}$}}}
\newcommand{\hattildeplane}[2]{\widetilde{\widehat{\mbox{O}{#1}}}^
           {\raisebox{-8.5pt}{$\scriptstyle {#2}$}}}

\newcommand{\be}{\begin{equation}}
\newcommand{\ee}{\end{equation}}
\newcommand{\bea}{\begin{eqnarray}}
\newcommand{\eea}{\end{eqnarray}}

\preprint{CALT-68-2321\\ CITUSC/01-007 \\{\tt hep-th/0103183}}

\title{Orientifolds, RR Torsion, and K-theory}

\author{Oren Bergman
\\
California Institute of Technology, Pasadena CA 91125, USA
\\ and\\
CIT/USC Center for Theoretical Physics \\
Univ. of Southern California, Los Angeles CA \\
\email{bergman@theory.caltech.edu}}
\author{Eric Gimon\\
School of Natural Sciences \\ Institute for Advanced Study \\
Einstein Drive, Princeton NJ 08540,
USA\\
\email{gimon@sns.ias.edu}}
\author{Shigeki Sugimoto\\
CIT/USC Center for Theoretical Physics \\
Univ. of Southern California, Los Angeles CA \\
\email{sugimoto@citusc.usc.edu}}

\abstract{We analyze the role of RR fluxes in orientifold
backgrounds from the point of view of K-theory, and demonstrate
some physical implications of describing these fluxes in K-theory
rather than cohomology. In particular, we show that certain fractional
shifts in RR charge quantization due to discrete RR fluxes are naturally 
explained in K-theory. We also show that some orientifold backgrounds,
which are considered distinct in the cohomology classification,
become equivalent in the K-theory description, while others become
unphysical.}

\begin{document}

\section{Introduction}

Orientifold planes are intriguing objects in many ways.
They appear, at least perturbatively, to be void of dynamics, 
and in this respect they are similar to orbifolds.
On the other hand orientifold planes are also similar to D-branes,
in that they appear with the same dimensionalities, carry RR charge, 
and have a non-vanishing tension.
However the latter two differ from their D-brane values in two important 
ways.
In some cases the tension of an orientifold plane is negative,
and in some cases its charge (and tension) is fractional.

There are actually a variety of orientifold planes 
for each dimensionality $p$, associated with the possibility of
turning on discrete fluxes of certain anti-symmetric tensor fields
\cite{Witten_Baryon,HK}.
These fluxes come in two varieties corresponding to the NSNS three-form
$H$, and the RR $(6-p)$-form $G_{6-p}$, and have 
traditionally been classified by the (integral) cohomologies
$H^3(\RP{8-p})$ and $H^{6-p}(\RP{8-p})$, respectively.
The space $\RP{8-p}$ is the ``sphere'' that surrounds the O$p$-plane
in the reduced space of the orientifold projection.
Except for the case $p=6$, both of these cohomologies are $\Z_2$,
so the fluxes are torsion classes. This means that there are four variants
of the orientifold plane for each $p$ (for $p\leq 5$),
depending on the values of the discrete fluxes (table~1).
For $p<2$ there are actually additional variants, which are associated 
with a discrete $G_{2-p}$-flux \cite{HK}. 
\begin{table}
\begin{center}
\begin{tabular}{|l|l|l|l|}
\hline
 $(H,G_{6-p})$ & O$p$ variant & D$p$ gauge group & charge  \\
\hline
 $(0,0)$ & O$p^-$ & $SO(2n)$ & $-2^{p-5}$ \\[5pt]
 $({1\over 2},0)$ & O$p^+$ & $USp(2n)$ & $+2^{p-5}$ \\[5pt]
 $(0,{1\over 2})$ & $\tildeplane{p}{-}$ & $SO(2n+1)$ &
  $-2^{p-5}+{1\over 2}$ \\[5pt]
 $({1\over 2},{1\over 2})$ &  $\tildeplane{p}{+}$ & $USp(2n)$ &
   $+2^{p-5}$ \\[3pt]
\hline
\end{tabular}
\caption{Orientifold plane variants and their charges.}
\end{center}
\end{table}

One immediately observes that orientifold planes with $p\leq 4$
are fractionally charged (relative to D$p$-branes).
This appears to be at odds with Dirac's quantization condition,
suitably generalized to higher rank anti-symmetric tensor fields,
or equivalently with the idea that RR charges and fields take values in
integral cohomology. 
In particular, since the O$p$-plane is charged
under $C_{p+1}$, the relevant cohomology group is given by
$H^{8-p}(\RP{8-p}) = \Z$. The generator of this group has a unit
D$p$-brane charge, so the O$p$-plane charge $\pm 2^{p-5}$ is not an
allowed value for $p\leq 4$. In addition, the presence of RR torsion
in $H^{6-p}(\RP{8-p})$ shifts the charge by 1/2 for the O$p^-$ plane,
but not for the O$p^+$ plane. So in the first case it seems to lead to
a further violation of Dirac's condition.

In some cases the violation can be attributed to an anomaly
in the path integral for the dual magnetic object, {\em i.e.}
a D$(6-p)$-brane, and can be expressed in terms of spacetime
fields \cite{Witten_anomaly}, but this is not generally the case.
For example, the $\pm 1/2$ charge of the O$4^\pm$ plane is consistent
with an anomalous phase in the D2-brane path integral coming from the
fermion determinant \cite{Witten_anomaly}. However
one needs another contribution to the phase in order to
explain the $0$ charge of the $\tildeplane{4}{-}$ variant,
and yet another to explain the $+1/2$ charge of $\tildeplane{4}{+}$.
One could imagine that the latter is due to the
term $\int B\wedge C_1$ in the D2-brane action, but there
is no obvious candidate for the former contribution.
For other orientifold planes one does not even have an analog
of the anomaly contribution of \cite{Witten_anomaly}.

All this indicates that integral cohomology is not the appropriate
mathematical framework for RR fields.  
We are familiar with the idea that D-branes, or more properly RR
charges, take values in K-theory rather than in integral cohomology
\cite{Moore_Minasian,Witten_K,Horava_K}.
This follows from the fact that a D$p$-brane is characterized not only
by its charge under the RR field $C_{p+1}$, but also by a gauge
bundle, which in turn determines its charges under lower rank RR
fields. It has recently been suggested that the RR fields take values
in K-theory as well \cite{Moore_Witten,Freed_Hopkins}. 
This is intuitively obvious
for RR fields created by D-brane sources, since the field at infinity
should belong to the same class as the source which creates it. But the
idea is that all RR fields, including source-free RR fluxes, are
classified in K-theory.

In this paper we investigate some of the consequences of describing
RR fluxes in K-theory rather than integral cohomology. 
In particular, this provides a partial explanation of the
observed fractional charges; K-theory correlates different
RR fields in such a way, that the discrete flux of $G_{6-p}$
alters the quantization of $G_{8-p}$-flux.
For $p=6$ the former ($G_0$) is integral rather than discrete, but
there is an analogous correlation between the parity of $G_0$
and the half-integral part of $G_2$.
Thus we are able to compute the shift in the quantization condition
of the RR charge due to the discrete RR flux. We are
not able to establish the quantization condition for the absolute
charges however.
In comparing the K-theory classification with the cohomological
one we also discover that some discrete fluxes are actually trivial,
while others are obstructed. This has interesting physical
implications concerning some of the lower orientifold planes.

The rest of the paper is organized as follows.
In section 2 we review what is known about orientifold planes,
and their classification in terms of discrete fluxes in cohomology.
In section 3 we describe the necessary K-theory machinery required
for describing RR fields in orientifold backgrounds, and compute
the relevant groups. We also explain how to compare the K-theory results
with cohomology using a real version of the Atiyah-Hirzebruch spectral
sequence (AHSS), which is further elaborated in the appendix.
In section 4 we discuss three implications of the difference between
K-theory and cohomology in orientifolds. The first concerns the question
of proper RR charge quantization, the second is the identification
in K-theory of certain orientifold variants which are distinct in 
cohomology, and the third is the dismissal of certain orientifold 
variants as unphysical. Section 5 is devoted to a separate discussion
of the orientifold 6-plane, which in many ways was the original motivation
for this work.

\section{Review of orientifolds}

\subsection{Basics}

Orientifolds are defined by gauging a discrete symmetry in string
theory which includes a reversal of the string
world-sheet orientation (denoted $\Omega$). The loci of points
which are fixed under this symmetry are called orientifold planes.
In particular, an orientifold $p$-plane
(O$p$) in Type II string theory is given by the $\Z_2$ projection
\be
\label{O_projection}
 \mbox{O}p:\qquad \R^{1,p}\times\R^{9-p}/{\cal I}_{9-p}\,\Omega\cdot
 \left\{
 \begin{array}{ll}
  1 & \;\; p=0,1\;\; \mbox{mod}\;\; 4 \\
  (-1)^{F_L} & \;\; p=2,3 \;\; \mbox{mod}\;\; 4\;,
 \end{array}\right.
 \ee
where $p$ is even in IIA and odd in IIB. The presence of
$(-1)^{F_L}$ for $p=2,3$ mod 4 can be understood from the
requirement that the generator square to 1 on fermions. For
example, $({\cal I}_2\Omega)^2=-1$ on fermions, but $({\cal
I}_2\Omega (-1)^{F_L})^2=({\cal I}_2\Omega)^2 (-1)^{F_L+F_R}=+1$.
The action of the orientifold on the antisymmetric tensor fields
is given by
\be
\begin{array}{lcll}
\label{field_action}
 B & \longrightarrow & -B & \\
 C_{p'} & \longrightarrow & +C_{p'} & \;\; p'=p+1\;\;
    \mbox{mod}\;\; 4 \\
 C_{p'} & \longrightarrow & -C_{p'} & \;\; p'=p+3\;\;
    \mbox{mod}\;\; 4\;.
\end{array}
\ee
In particular, the projection leaves the D$p$-brane invariant,
and in fact preserves the same 16 supersymmetries.

If we include $n$ D$p$-branes (and their images),
there are two possible actions
on the open string Chan-Paton factors \cite{GP},
\be
 \lambda\rightarrow M^{-1}\lambda^T M \;, \quad
 M=\I_{2n\times 2n} \;\;\mbox{or}\;\;
 \left[
 \begin{array}{cc}
  0 & i\I_{n\times n} \\
  -i\I_{n\times n} & 0
 \end{array}
 \right] (\equiv \J) \;.
\ee
In the first case the resulting gauge group is $SO(2n)$, and in the second
case $USp(2n)$. The corresponding M\"obius strip amplitudes, and therefore
also the vacuum $\RP{2}$ amplitudes,  have the opposite sign,
so we see that there are actually two types of O$p$-planes.
These are denoted O$p^-$ and O$p^+$ for the $SO(2n)$ and $USp(2n)$
cases, respectively. Comparing the M\"obius strip amplitude
with the cylinder amplitude, one finds that the RR charge and tension
of the O$p$-planes (as compared with those of the D$p$-branes) is
given by $-2^{p-5}$ for O$p^-$, and $+2^{p-5}$ for O$p^+$
(hence the sign in the notation).

\subsection{Cohomology torsion variants}

Another way to see that there are (at least) two variants
(at least for $p\leq 6$)
is to note that the orientifold background admits a discrete degree
of freedom corresponding to the holonomy of the $B$ field
\cite{Witten_Baryon},
\be
\label{B-holonomy}
 b = \int_{\RP{2}} {B\over 2\pi} = 0 \;\; \mbox{or} \;\; {1\over 2}\;,
\ee
where $\RP{2}\subset\RP{8-p}$ which surrounds the O$p$-plane.
A non-trivial holonomy contributes $e^{2\pi i b} = -1$ in the
$\RP{2}$ amplitude, and therefore exchanges the O$p^-$ and
O$p^+$ planes.\footnote{It isn't obvious at this stage which variant
should be identified with trivial holonomy, and which with non-trivial
holonomy. This will be clarified in K-theory.}
Since $B$ is odd under the projection, which maps opposite points
in the double cover of $\RP{8-p}$, the holonomy
is an element of $H^2(\RP{8-p},\widetilde{U(1)})$,
where $\widetilde{U(1)}$ is the twisted $U(1)$ bundle.
This has
the same torsion subgroup as $H^3(\RP{8-p},\widetilde{\Z})$
($\widetilde{\Z}$ being the twisted $\Z$ bundle),
so we can identify the field strength $H=dB$ as an element
of the latter,
\be
\label{H-torsion}
 [H] \in H^3(\RP{8-p},\widetilde{\Z}) = \Z_2 \;,
\ee
where the choice of torsion coincides with the holonomy $b$.

RR fields can also have a discrete holonomy, which leads to
additional variants of the orientifold plane. For the O$p$-plane
with $p\leq 5$
 the relevant holonomy is
\be
\label{C-holonomy}
  c = \int_{\RP{5-p}} {C_{5-p}\over 2\pi} = 0 \;\; \mbox{or} \;\; {1\over 2}\;,
\ee
and is identified with a torsion $G_{6-p}$-field,
\be
\label{G-torsion}
 [G_{6-p}] \in \left\{
 \begin{array}{ll}
  H^{6-p}(\RP{8-p},\Z) & \;\; \mbox{for}\;\; p=0,2,4 \\
  H^{6-p}(\RP{8-p},\widetilde{\Z}) & \;\; \mbox{for}\;\;p=1,3,5
 \end{array}\right\}
 = \Z_2 \;.
\ee 
The choice of a non-trivial $c$ is denoted
$\tildeplane{p}{\pm}$.
Unlike the previous case, the charges of these variants cannot be
computed in perturbation theory, since they involve a RR background.
Instead, one must appeal to arguments involving strong coupling. For
example, the O3-plane has four variants corresponding to $H$ and
$G_3$ torsion, both of which take values in
$H^3(\RP{5},\widetilde{\Z})=\Z_2$. S-duality in Type IIB string theory
exchanges $H$ and $G_3$, so O$3^-$ and $\tildeplane{3}{+}$
are invariant, and O$3^+$ is interchanged with
$\tildeplane{3}{-}$  \cite{Witten_Baryon}. If we include
D3-branes, string S-duality becomes Olive-Montonen duality in the
world-volume $N=4$ supersymmetric gauge theory. The $SO(2n)$ theory
corresponding to O$3^-$ is invariant under this duality, but the
$USp(2n)$ theory of O$3^+$ is related to an $SO(2n+1)$ theory
\cite{Elizur}. We can therefore interpret the $\tildeplane{3}{-}$
plane as an O$3^-$ plane with an additional fractional (1/2) D3-brane.
The fourth variant, $\tildeplane{3}{+}$, corresponds to turning
on $H$ torsion on $\tildeplane{3}{-}$. This exchanges the
orthogonal gauge group with a symplectic one, $USp(2n)$. The theory is
perturbatively the same as for O$3^+$, but the BPS dyon spectrum is
different \cite{HK}. This $USp(2n)$ theory is self-dual.

There are analogous strong coupling arguments for Type IIA O2 planes
\cite{Sethi} and O4 planes \cite{Gimon,Hori,Lowe} using M-theory.
However, it is clear from T-duality that this pattern persists for all
$p\leq 5$ (see table~1).

The case $p=6$ is special, since
$[G_0]\in H^0(\RP{8-p},\Z)=\Z$, rather than $\Z_2$. 
These classes correspond to
the allowed (integral) values of the cosmological constant in
massive Type IIA supergravity \cite{Romans,Polchinski}.
It does not appear, at this stage, to label distinct O$6$-planes,
but rather different IIA backgrounds in which an O$6$-plane can sit.
One might therefore conclude that the O$6$-plane only has two
variants, O$6^-$ and O$6^+$. However, we shall see in section~5
that there is another variant, which has the properties (namely charge
and tension) of the $\tildeplane{p}{-}$-planes, and is therefore
similarly denoted $\tildeplane{6}{-}$ (see also \cite{Sugimoto}).

For $p=0$ and 1 there is another possible holonomy given by
\be
\label{C'-holonomy}
  c' = \int_{\RP{1-p}} {C_{1-p}\over 2\pi} = 0 \;\; \mbox{or} \;\; 
{1\over 2}\;.
\ee 
This is related to the torsion groups
$H^1(\RP{7},\widetilde{\Z})$ and $H^2(\RP{8},\Z)$. We denote the
corresponding variants $\widehat{\mbox{O}1}$ and
$\widehat{\mbox{O}0}$. When both RR torsions are turned on we get
more variants, $\widehat{\widetilde{{\mbox{O}1}}}$ and
$\widehat{\widetilde{{\mbox{O}0}}}$.
The charges of these variants were not known.
Using K-theory, we will be able to determine their charges.
We will also see
that some of these variants are removed when one interprets RR
fields in K-theory. For the O2-plane, one could also consider
a $G_0$ background, which takes values in $H^0(\RP{6},\Z)=\Z$.
The situation is similar to the O6-plane case above,
and will also be addressed in section 5.

\subsection{Brane realization of torsion}

Pure torsion fields can be realized physically by simple brane
constructions \cite{EJS,Johnson,HK}. Consider an O$p$-plane ($p\leq 5$)
situated along
the directions $\{x^1,\ldots,x^{p-1},x^6\}$, and either an
NS5-brane along $\{x^1,\ldots,x^5\}$ for the case of the $H$
field, or a D$(p+2)$-brane along
$\{x^1,\ldots,x^{p-1},x^7,x^8,x^9\}$ for the case of the
$G_{6-p}$ field (figure 1).\footnote{A similar construction can be
made for the additional RR torsion ($c'$)
in the case of $p=0,1$, using an intersection with a D6-brane
and D7-brane, respectively.} In both cases the brane intersects
the orientifold plane along $\{x^1,\ldots,x^{p-1}\}$, and
therefore forms a ``domain wall'' on the O$p$-plane. The brane is
``fractional'' in the sense that it is its own image under the
orientifold projection, and can therefore not move off the
orientifold plane. It is therefore a source of 1/2 unit of flux in
the reduced space, i.e. 
\be
 \int_{\SSS^{3,1}} {H\over 2\pi} = {1\over 2}
\ee
for the NS5-brane, and
\be
 \int_{\SSS^{6-p,1}} {G_{6-p}\over 2\pi} = {1\over 2}
\ee
for the D$(p+2)$-brane.\footnote{We use $\SSS^{l,m}$ to denote the
sphere in the reduced space $\R^{l,m}\equiv \R^{l+m}/\Z_2$,
where the $\Z_2$ inverts $l$ of
the coordinates. In particular $\SSS^{l,0}\sim\RP{l-1}$.}
Now remove two small disks at the intersection of the sphere
with the orientifold plane, and apply Stoke's law to get
\be
 \Delta \int_{\RP{2}} {B\over 2\pi} = 
 \Delta \int_{\RP{5-p}} {C_{5-p}\over 2\pi} = {1\over 2}
\ee
for the difference in the holonomy between the two sides.
The field strengths $H$ and $G_{6-p}$ are trivial (as differential
forms) at infinity, but the above result shows that there is relative
torsion between the two pieces of the O$p$-plane.
The NS5-brane interchanges the O$p^-$ and O$p^+$ planes,
and the D$(p+2)$-brane interchanges the O$p$ and and
$\widetilde{\mbox{O}p}$ planes.
\begin{figure}
\label{brane_torsion}
\centerline{\epsfxsize=2in\epsfbox{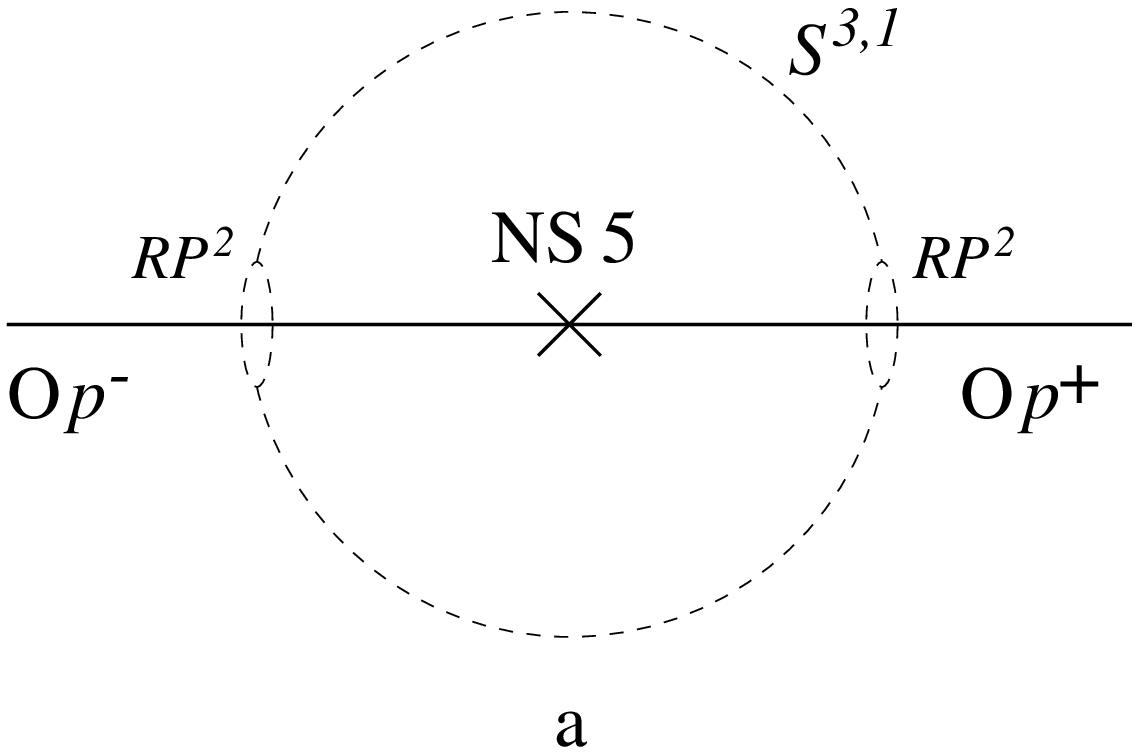}\hspace{1cm}
\epsfxsize=2in\epsfbox{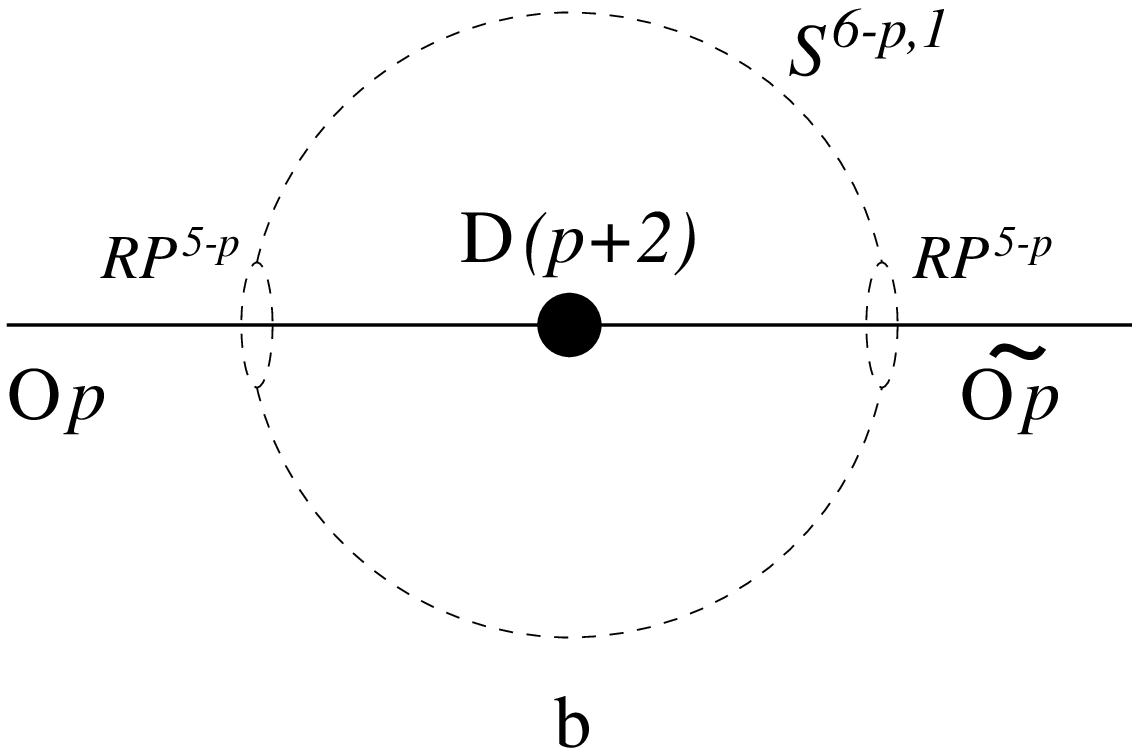}}
\medskip
\caption{Brane realization of (a) $H$ torsion, (b) $G_{6-p}$ torsion.}
\end{figure}

For two coincident branes the flux is 1, but they are now
free to move off the orientifold plane (in opposite directions).
The torsion part of the field is therefore trivial, and the orientifold
plane remains unchanged. 

For $p=6$ the D-brane picture (figure 1b) is
special, in that $G_0$ is created by a D8-brane, which fills
the entire transverse space. The D8-branes can therefore never move off
the orientifold plane, and this is consistent with the fact that $G_0$
is integral rather than torsion.

We can try to use this picture to understand the relative charges
of the $\tildeplane{p}{\pm}$ planes.
Consider a ``deformation'' of the intersecting D-brane configuration,
in which the D$(p+2)$-brane wraps
$\R^p\times \RP{2}$ at $x^6\rightarrow\infty$, and ``opens up'' into
a flat $\R^{p+2}$ at $x^6=0$ (figure~2). The $\RP{2}$ is not a homology
cycle in the bulk, so the wrapped part of the brane shrinks to zero size,
giving the original intersection configuration.
If we assume that the world-volume gauge field on the D$(p+2)$-brane
satisfies \cite{Aharony,Sugimoto}
\be
\label{Dirac_shift}
 \int_{\RP{2}}{F\over 2\pi} = {1\over 2} \;\;\mbox{mod}\;\;\Z\;,
\ee
then the charge of the O$p^-$ plane would shift by 1/2,
due to the coupling $\int F\wedge C_{p+1}$. 
On the other hand the charge of O$p^+$ could only change by an integral
amount due the additional coupling $\int B\wedge C_{p+1}$, 
since the $B$ field also has 1/2 unit of flux on $\RP{2}$.
This would therefore explain the observed relative charges
of these two variants. A result similar to (\ref{Dirac_shift}) was
obtained in \cite{Freed_Witten} for orientable manifolds for which
$w_2$ is non-trivial. However $\RP{2}$ is non-orientable, so the question
remains open.
\begin{figure}[h]
\label{wrapped_brane}
\centerline{\epsfxsize=4in\epsfbox{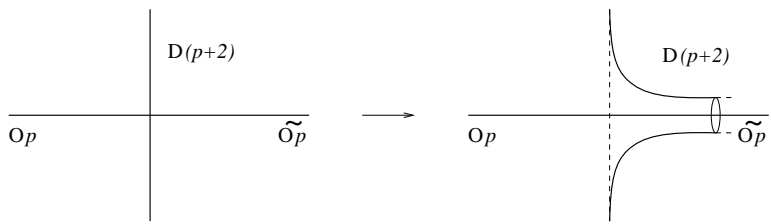}}
\medskip
\caption{Deforming the intersecting brane to a wrapped brane.}
\end{figure}

\section{Orientifolds in K-theory}

In ordinary (non-orientifold) Type IIA and IIB backgrounds
RR fluxes take values in $K(X)$ and $K^{-1}(X)$,
respectively \cite{Moore_Witten}. This complements the fact that RR
charges, or D-branes, are classified by $K^{-1}(X)$ and $K(X)$,
respectively.\footnote{If $X$ is non-compact, then D-branes are
actually classified by K-theory with compact support,
$K^{-1}_{cpct}(X)$ and $K_{cpct}(X)$, or more precisely by the kernel
of the natural maps from these groups to the ones without compact
support.} One way to think about this, which will be useful in
generalizing to orientifolds, is to divide the space into two regions
using a domain wall made of D8-branes and anti-D8-branes in IIA, and
non-BPS D8-branes in IIB. Differences in the values of the RR fluxes
between the two regions are then in one-to-one correspondence with the
RR charges carried by the D8-brane system, and are therefore
classified by $K(X)$ in IIA, and by $K^{-1}(X)$ in IIB. The groups are
exchanged relative to the D-brane classification since there we start
with a D9-$\overline{\mbox{D9}}$ system in IIB, and non-BPS D9-branes
in IIA.

\subsection{Real K-theory}

The orientifold backgrounds of interest are defined by a $\Z_2$
projection on closed strings that includes world-sheet parity $\Omega$,
and an action on the open string Chan-Paton factors that squares to
either 1 or $-1$. In K-theoretic terms, one is given an involution
$\tau$ on the space $X$, and an antilinear involution on the bundles
that commutes with $\tau$. This defines the real and quaternionic
K-theory groups $KR(X)$ and $KH(X)$, depending on whether the
involution on the bundles squares to 1 or $-1$ \cite{Atiyah}. Higher
groups can then be defined by (assuming compact support)
\be
\label{higher}
 KR^{p,q}(X) = KR(X\times \R^{p,q})\;,
\ee
where the involution acts on $p$ coordinates in $\R^{p,q}$,
and similarly for $KH$. These satisfy the periodicity conditions
\begin{eqnarray}
\label{Bott}
 KR^{p,q}(X) &=& KR^{p+1,q+1}(X) \\
 KR^{p,q}(X) &=& KR^{p+8,q}(X) \;,
 \end{eqnarray}
and similarly for $KH$.
The first implies that $KR^{p,q}$ depends only on the difference $p-q$,
so we denote $KR^{p,q}(X)=KR^{p-q}(X)$.
In addition $KR$ and $KH$ are related by
\be
\label{KRKH}
 KR^{-n}(X) = KH^{-n+4}(X) \;.
\ee
Another variant of real K-theory can be defined by considering
an involution that also exchanges the two bundles $E$ and $F$.
Physically this corresponds to the inclusion of the operator
$(-1)^{F_L}$ ({\em e.g.} for $p=3$ mod 4 in (\ref{O_projection})),
which exchanges the branes and antibranes.
The corresponding groups are denoted $KR_\pm(X)$ and $KH_\pm(X)$,
and higher versions are defined precisely as in (\ref{higher}).
These groups are related to ordinary real K-theory as follows,
\be
\label{hopkins}
 KR_\pm(X) = KR(X\times \R^{0,2}) = KR^{-2}(X) \;,
\ee
and similarly for $KH_\pm$.\footnote{This relation is the analog in real
K-theory of Hopkins' relation in complex equivariant K-theory,
$K_\pm(X) = K_{\Z_2}(X\times \R^{1,1})$.
The latter can be understood as the failure of ordinary Bott periodicity
when one tries to construct the Dirac index map:
$K_{\Z_2}(X\times \R^{1,1})\rightarrow K_{\Z_2}(X)$.
Orientation reversal on the $\R^2$ exchanges the
two spinor chiralities, and therefore gives $K_\pm(X)$ rather
than $K_{\Z_2}(X)$. In the real case it's exactly the opposite, since
complex conjugation also exchanges the two chiralities.
Bott periodicity works for $\R^{1,1}$ (\ref{Bott}), and fails
for $\R^{0,2}$ and $\R^{2,0}$. The former gives $KR_\pm(X)$,
and the latter $KH_\pm(X)$.}

\subsection{RR fields in orientifold backgrounds}

To classify RR fields in the background of an orientifold
$p$-plane we divide the transverse space $\R^{9-p}$ by wrapping D8-branes
on the transverse sphere $\SSS^{8-p}$. In IIA we again consider
both D8-branes and anti-D8-branes, and in IIB we use the non-BPS
D8-branes. The orientifold projection (\ref{O_projection})
acts freely on the sphere, giving $\RP{8-p}$, and the corresponding
K-groups are obtained by keeping track
of how the projection acts on the D8-branes and their bundles.

For example, the O$0$ projection $\R^9/{\cal I}_9\,\Omega$ in IIA
maps a D8-brane wrapping $\SSS^8$ with a given orientation to a  
D8-brane wrapping it with the opposite orientation. It therefore 
exchanges wrapped D8-branes with wrapped anti-D8-branes, and vice-versa.
The group is therefore either
$KR_\pm(\SSS^{9,0})$ or $KH_\pm(\SSS^{9,0})$, depending on whether the
action on the Chan-Paton bundle squares to 1 or $-1$.
Following the general arguments of \cite{GP}, we associate the former
to the O$0^-$-plane, and the latter to the O$0^+$-plane.

The usual 8-fold periodicity of orientifolds (and real K-theory)
implies that for the O$8^-$ and O$8^+$ planes the groups are also
$KR_\pm$ and $KH_\pm$, respectively. This is somewhat counterintuitive,
in that a D8-brane gets mapped to an image D8-brane and not an
anti-D8-brane.
However recall that the D8-branes wrap the transverse sphere, which
in this case is just two points on either side of the O8-plane.
This means that the orientation of the ``wrapped'' D8-brane is reversed
on the other side. The projection maps the D8-brane to a D8-brane
of the same orientation on the other side, so it looks like a ``wrapped''
anti-D8-brane.

The appropriate groups for the other orientifold planes can be
similarly determined. The results are summarized in
table~2.
\begin{table}[htb]
\centerline{
\begin{tabular}{|l|l|}
\hline
 $Op^-$ & K-group  \\ \hline\hline
 $O0^-$ & $KR_{\pm}(S^{9,0}) = \Z \oplus \Z_2$ \\
 $O1^-$ & $KR^{-1}(S^{8,0}) = \Z \oplus \Z_2$ \\
 $O2^-$ & $KR(S^{7,0}) = \Z \oplus \Z$ \\
 $O3^-$ & $KH^{-1}_{\pm}(S^{6,0})= \Z$ \\
 $O4^-$ & $KH_{\pm}(S^{5,0}) = \Z$ \\
 $O5^-$ & $KH^{-1}(S^{4,0}) = \Z$ \\
 $O6^-$ & $KH(S^{3,0}) = \Z \oplus \Z$ \\
 $O7^-$ & $KR^{-1}_{\pm}(S^{2,0}) = \Z$ \\
 $O8^-$ & $KR_{\pm}(S^{1,0}) = \Z$  \\
\hline
\end{tabular}
\hspace{1cm}
\begin{tabular}{|l|l|}
\hline
 $Op^+$ & K-group  \\ \hline\hline
 $O0^+$ & $KH_{\pm}(S^{9,0}) = \Z$ \\
 $O1^+$ & $KH^{-1}(S^{8,0}) = \Z$ \\
 $O2^+$ & $KH(S^{7,0}) = \Z \oplus \Z$ \\
 $O3^+$ & $KR^{-1}_{\pm}(S^{6,0})= \Z$ \\
 $O4^+$ & $KR_{\pm}(S^{5,0}) = \Z\oplus\Z_2$ \\
 $O5^+$ & $KR^{-1}(S^{4,0}) = \Z\oplus\Z_2$ \\
 $O6^+$ & $KR(S^{3,0}) = \Z \oplus \Z$ \\
 $O7^+$ & $KH^{-1}_{\pm}(S^{2,0}) = \Z$ \\
 $O8^+$ & $KH_{\pm}(S^{1,0}) = \Z$  \\
\hline
\end{tabular}
}
\caption{\small Orientifold K-theory groups for RR fields.} \label{t1}
\end{table}
Using (\ref{Bott})-(\ref{hopkins}) we can re-express these groups as
$KR^{p-10}(\SSS^{9-p,0})$ for O$p^-$, and
$KH^{p-10}(\SSS^{9-p,0})=KR^{p-6}(\SSS^{9-p,0})$ for O$p^+$.
Note that, as expected, the groups which classify RR fields are
shifted by one relative to the groups which classify RR charges,
$KR^{p-9}$ and $KH^{p-9}$ \cite{Gukov,Hori_K,BGH}.
For example, D-branes in an O8 background (Type IA) are
classified by $KR^{-1}$, whereas RR fields take values in
$KR_\pm=KR^{-2}$. To compute the groups we used the following decomposition
isomorphism due to Atiyah \cite{Atiyah}
\be
 KR^{-n}(\SSS^{m,0}) = KR^{m-n+1}(pt.) \oplus KR^{-n}(pt.) 
\qquad (m\geq 3)
\;,
\ee
and our knowlege of the real K-groups of a point,
\be
 KR^{-m}(pt.) = \{\Z,\Z_2,\Z_2,0,\Z,0,0,0\} \;\; \mbox{mod} \;\; 8 \;.
\ee


\subsection{Comparing with cohomology}

The above results agree with cohomology in some cases,
and differ in other cases. The free ($\Z$) part of K-theory always
agrees with cohomology. This corresponds to the RR flux of the
orientifold $p$-plane itself, {\em i.e.} $G_{8-p}$.
For $p=2,6$ there is an additional free part corresponding to
$G_0$, which also agrees with cohomology.
The difference between K-theory and cohomology can only be in
the torsion subgroup.\footnote{This is a consequence of the Chern
isomorphism, {\em e.g.}
$K(X)\times\QQ \simeq H^{even}(X,\QQ)$.
In the real case the cohomologies will be twisted or normal, according
to the action of the orientifold on the corresponding RR field
(\ref{field_action}).}

For example, the relevant cohomologies for
orientifold 5-planes are
\be
\label{O5_cohomology}
 H^3(\RP{3},\Z)\oplus H^1(\RP{3},\widetilde{\Z}) = \Z \oplus \Z_2 \;.
\ee The first term corresponds to the integral $G_3$ flux of
the O5-plane, and the second is $G_1$ torsion, which
gives the $\widetilde{\mbox{O}5}$ variants. Let us once again
stress that these cohomologies do not incorporate the fractional
shift in the $G_3$ flux of $\tildeplane{5}{-}$ relative
to O$5^-$. In K-theory we find \be
\begin{array}{lcll}
\label{O5_K}
 KH^{-1}(\SSS^{4,0}) & = & \Z & \;\; \mbox{for O$5^-$} \\
 KR^{-1}(\SSS^{4,0}) &= & \Z\oplus \Z_2 & \;\; \mbox{for O$5^+$} \;.
\end{array}
\ee How should we interpret the absence of the torsion subgroup
in the first case?
Clearly we need to understand better how to relate (or ``lift'')
cohomology to K-theory.

There is in fact a systematic algorithm which does exactly this,
known as the Atiyah-Hirzebruch spectral sequence 
(AHSS).\footnote{Introductory reviews of the AHSS for complex
K-theory can be found in the appendix, as well as in the appendices of 
\cite{DMW} and \cite{BGK}.} 
The basic idea of the AHSS is to compute $K(X)$ using
a sequence of successive approximations, starting with integral
cohomology $H^*(X,\Z)$. Each successive step is given by the
cohomology of a differential $d_r$ that acts on classes in the
previous step. Thus classes which are not $d_r$-closed are removed
in the new approximation, and classes which are $d_r$-exact are
trivial. For example, $d_1$ is just the usual co-boundary operator
acting on co-chains (which are the zero'th step), so the first
step is ordinary de-Rahm cohomology. Higher differentials, if they
are non-trivial, will refine the approximation by
removing some cohomology classes, and trivializing others.
The first non-trivial higher differential in the complex case
is given by $d_3=Sq^3+H$, where $Sq^3$ is the Steenrod square
\cite{DMW}.
The only other differential which may be non-trivial in ten dimensions
is $d_5$.

The AHSS can be extended to real K-theory as well. In the appendix
we describe how to do this in the case of a freely acting
involution, such as the ones appearing in table~2.
In this case the AHSS follows a similar pattern to the complex
case, with the modification that
the cohomologies appearing in the first approximation can be
twisted or normal, in accordance with the orientifold action on
the RR fields (\ref{field_action}). Correspondingly, the
differential $d_3$ maps between ordinary and twisted cohomologies,
and must therefore itself be twisted. The precise form of $d_3$ is
not known, but it can presumably be expressed in the form
$d_3=\widetilde{Sq^3}+H$, where $\widetilde{Sq^3}$ is a twisted
version of the Steenrod square. Here the second term is torsion,
since $[H]\in H^3(\RP{8-p},\widetilde{\Z}) = \Z_2$.
As we have seen before, this provides
the difference between the O$p^-$-plane and the
O$p^+$-plane. We shall assume that the first term is trivial in
both cases, and justify this assumption is hindsight.
We shall also see that $d_5$ (which is untwisted) is trivial in all cases.

Actually, what one really computes in the AHSS is an ``associated
graded complex'' of $K(X)$ (or any of its variants),
denoted $\mbox{Gr}K(X)$. This is
defined in terms of a filtration of $K(X)$, $K_n(X)\subseteq
K_{n-1}(X) \subseteq\cdots\subseteq K_0(X)=K(X)$, as \be
\label{GrK}
  \mbox{Gr}K(X) = \oplus_p K_p(X)/K_{p+1}(X) \;,
\ee where $K_p(X)$ is the subgroup of classes in $K(X)$ which are
trivial on the $(p-1)$-tree of $X$. The successive ratios
appearing in (\ref{GrK}) are precisely the objects one computes by
the AHSS. In particular, in the first order approximation
\be
 K_p(X)/K_{p+1}(X) = \left\{
 \begin{array}{ll}
 H^p(X,\Z) & \;\;\mbox{for}\;\;p\;\;\mbox{even}\\
 0 & \;\;\mbox{for}\;\;p\;\;\mbox{odd}\;.
 \end{array}\right.
\ee
To get the actual subgroups $K_p(X)$, and in particular
$K(X)$ itself, requires solving extension problems of the form
\be
\label{extension}
 0\rightarrow K_{p+1}(X) \rightarrow
 K_{p}(X) \rightarrow
 K_{p}(X)/K_{p+1}(X) \rightarrow 0 \;.
\ee
If the above exact sequence splits, {\em i.e.} if it is exact
both ways, the extension is said to be trivial, and
$K_{p}(X)=K_{p+1}(X)\oplus K_{p}(X)/K_{p+1}(X)$. If all the
extensions are trivial we deduce recursively that
$K(X)=\mbox{Gr}K(X)$.
In other words, K-theory is isomorphic to the ring of cohomologies
that survive the higher differentials of the AHSS.
On the other hand, if some of the extensions are non-trivial
$K(X)$ will differ from its graded complex.
In particular, the additive structure in K-theory will be
different from that of the cohomology classes.
Physically this means that RR fields of different degree become correlated.

Consider for example the space $X=\RP{5}$, for which
\begin{eqnarray}
 \mbox{Gr}K(\RP{5}) &=& H^0(\RP{5}) \oplus H^2(\RP{5})
  \oplus H^4(\RP{5}) \nonumber \\
 &=&  \Z\oplus\Z_2\oplus\Z_2 \;.
\end{eqnarray}
The relevant extension is given by
\be
\begin{array}{ccccccccc}
 0 & \rightarrow & K_{3}(\RP{5}) & \rightarrow & K_{2}(\RP{5}) &
  \rightarrow & K_{2}(\RP{5})/K_{3}(\RP{5}) & \rightarrow  & 0 \\
 && \parallel && \parallel && \parallel && \\
 && H^4=\Z_2 && K_1(\RP{5}) && H^2=\Z_2 &&
\end{array} \;.
\ee
On the other hand it is known
(see for example the appendix of \cite{BGK}) that
\be
 K_1(\RP{5}) = \widetilde{K}(\RP{5}) = \Z_4 \;,
\ee
so this extension is non-trivial. The RR 2-torsion and 4-torsion
appearing in $H^2(\RP{5})$ and $H^4(\RP{5})$ become correlated in
K-theory; twice the generator of the former corresponds to the
generator of the latter.

In practical terms, one often needs additional information to
determine whether an extension is trivial or not, which precludes
the AHSS as a useful method for computing $K(X)$ (as opposed to
$\mbox{Gr}K(X)$). However it will be extremely useful in our case,
since we already know the precise groups (table~2). We are
therefore able to determine all the extensions (\ref{extension})
unambiguously, and thereby compare the K-theory data with
cohomology.

\section{Applications}

Having elaborated the K-theory machinery needed to classify RR fields
surrounding orientifolds, we would now like to demonstrate three
important implications of the fact that that RR fields are valued
in K-theory rather than cohomology.

\subsection{Charge quantization}

As was previously stated, there are two separate issues related
to the question of proper RR charge quantization. The first has to do
with the absolute fractional charges (of the O$p$-planes with $p\leq 4$,
for example), and the second with the relative charge induced by RR
torsion (1/2 for O$p^-$, 0 for O$p^+$).
The first issue encompasses the second, and should in principle
be resolved by an analog of the relation \cite{Moore_Witten}
\be
 {G(x)\over 2\pi} = \sqrt{\widehat{A}}\,\mbox{ch}(x) \;\; ,
 \;\; x\in K(X)
\ee
for real K-theory.
Unfortunately the generalization is not known, so we will concentrate
on the issue of the relative charge.
The main idea, as alluded to in the previous section, is that in
K-theory RR fields of different degree become correlated via non-trivial
extensions.

Note first that for the O8 and O7-planes K-theory is in complete
agreement with cohomology. There is only a single RR flux,
and therefore no possibility for a non-trivial extension.
There are only two variants of these planes, O$p^-$ and O$p^+$.
In particular the O8-plane (O7-plane) cannot support a fractional
D8-brane (D7-brane).
This is consistent with other arguments which rule out the
possibility of a gauge group $SO(2n+1)$ in these cases \cite{Sugimoto}.

The O6-plane background supports two RR fluxes, $G_2$ and $G_0$,
but both are integral. Since the differential $d_3$ (and $d_5$) is
obviously trivial
in this case,
the AHSS terminates at the first approximation, {\em i.e.}
\be
\begin{array}{rcl}
 K_0/K_1 &=& H^0(\RP{2},\Z) = \Z \nonumber \\
 K_1/K_2 &=& 0 \nonumber \\
 K_2 &=& H^2(\RP{2},\widetilde{\Z}) = \Z \;.
\end{array}
\ee
Using $K_0=KH(\SSS^{3,0})$ and $KR(\SSS^{3,0})$ for O$6^-$ and
O$6^+$, respectively, we can determine the filter groups uniquely,
\be
\begin{array}{l}
K_0 = \Z \oplus \Z \\
K_1=K_2=\Z \;.
\end{array}
\ee
However the extension given by
\be
\begin{array}{ccccccccc}
 0 & \rightarrow & K_1 & \rightarrow & K_0 & \rightarrow &
  K_0/K_1 & \rightarrow & 0  \\
 && \parallel && \parallel  && \parallel && \\
 && H^2=\Z && \Z\oplus \Z  && H^0=\Z &&
\end{array}
\ee
does not have a unique solution, so one cannot determine in this way
whether and how the two RR fluxes are correlated.
We will return to the O6-plane in section~5.

The first case where this procedure gives a non-trivial result
is the O$5$-plane. The cohomologies and K-theory for the O$5$-planes
were given in (\ref{O5_cohomology}) and (\ref{O5_K}).
The only possible non-triviality of $d_3$ could be in the map
\be
 H^0(\RP{3},\widetilde{\Z}) \stackrel{d_3}{\longrightarrow}
H^3(\RP{3},\Z) \;.
\ee
However $H^0(\RP{3},\widetilde{\Z})=0$, so $d_3$ (and $d_5$)
is trivial for both
O$5^-$ and O$5^+$. The AHSS again terminates at cohomology, and we find
\be
\begin{array}{lclcl}
 K_0 &=& K_1 &=& \left\{
 \begin{array}{ll}
  KH^{-1}(\SSS^{4,0}) = \Z & \;\;\mbox{for O}5^- \\
  KR^{-1}(\SSS^{4,0}) = \Z\oplus\Z_2 & \;\; \mbox{for O}5^+ \\
\end{array}\right. \\[15pt]
  K_2 &=& K_3 &=& H^3(\RP{3},\Z) = \Z \qquad\qquad \mbox {for both}\;.
\end{array}
\ee
This time we find that the extension given by
\be
\begin{array}{ccccccccc}
 0 & \rightarrow & K_2 & \rightarrow & K_1 & \rightarrow &
  K_1/K_2 & \rightarrow & 0  \\
 && \parallel &&  && \parallel && \\
 && H^3=\Z &&  && H^1=\Z_2 &&
\end{array}
\ee 
is trivial for O$5^+$, but non-trivial for O$5^-$.
The generator of $H^3$ maps to the generator of (the free part of)
$K_1$ for O$5^+$, but to twice the generator for O$5^-$. In the
latter case the generator of $K_1$ maps to non-trivial torsion in
$H^1$. We therefore see that $G_1$ torsion produces a
half-integral shift in the flux of $G_3$ for the O$5^-$ plane,
and has no effect on the charge of the O$5^+$ plane.

The story repeats along similar lines for the lower O$p^-$ planes.
The relevant cohomologies are $H^{8-p}(\RP{8-p})=\Z$ and
$H^{6-p}(\RP{8-p})=\Z_2$, which are appropriately twisted or untwisted.
The differential $d_3$ is trivial since $H=0$ (and we assume
$\widetilde{Sq^3}=0$). Possible non-trivialities of $d_5$ could arise
from the (twist preserving) maps
\begin{eqnarray}
 H^{3-p}(\RP{8-p}) & \longrightarrow & H^{8-p}(\RP{8-p}) \nonumber \\
 H^{1-p}(\RP{8-p}) & \longrightarrow & H^{6-p}(\RP{8-p}) \\
 H^{6-p}(\RP{8-p}) & \longrightarrow & H^{11-p}(\RP{8-p})\;, \nonumber
\end{eqnarray}
however in all cases the maps are found to be trivial.
The relevant extension is
\be
\begin{array}{ccccccccc}
 0 & \rightarrow & K_{7-p} & \rightarrow & K_{6-p} & \rightarrow &
  K_{6-p}/K_{7-p} & \rightarrow & 0  \\
 && \parallel && \parallel && \parallel && \\
 && H^{8-p}=\Z && \Z && H^{6-p}=\Z_2 &&
\end{array} \;,
\ee so the half-integral part of the $G_{8-p}$ flux, {\em i.e.}
the charge of the orientifold plane, is determined by the
$G_{6-p}$ torsion.

Actually, for $p\leq 2$ there is an additional RR cohomology $H^{2-p}$,
{\em i.e.}
\be
\label{moreRR}
\begin{array}{lcll}
 H^0(\RP{6},\Z) &=& \Z & \;\;\mbox{for}\;\; p=2 \\
 H^1(\RP{7},\widetilde{\Z}) &=& \Z_2 & \;\;\mbox{for}\;\; p=1 \\
 H^2(\RP{8},\Z) &=& \Z_2 & \;\;\mbox{for}\;\; p=0 \;.
\end{array}
\ee
As with the other cohomologies, these lift unobstructed to the graded
complex; $d_3$ is trivial since $H=0$ for O$p^-$ planes, and $d_5$
is trivial since $H^{7-p}(\RP{8-p})$ (appropriately normal or twisted)
is trivial. There is also an additional extension to solve,
\be
\begin{array}{ccccccccc}
 0 & \rightarrow & K_{3-p} & \rightarrow & K_{2-p} & \rightarrow &
  K_{2-p}/K_{3-p} & \rightarrow & 0  \\
 &&  &&  && \parallel && \\
 && && && H^{2-p} &&
\end{array} \;.
\ee
 One can show that in all three cases $K_{3-p}=\Z$, and
$K_{2-p}=K_0$, which is $\Z\oplus\Z$ for $p=2$, and $\Z\oplus\Z_2$
for $p=0,1$. The O$2^-$ case is similar to the O$6^-$ case above;
there is no unique solution to the extension, so one must use a
different method to determine the effect of $G_0$. We will
return to this problem in section~5. For O$1^-$ and O$0^-$ the
extension is trivial, so $G_{2-p}$ torsion does not shift their
charge.\footnote{This is also consistent with the fact that the
branes which create this torsion, {\em i.e.} a D6-brane for O$0^-$
and a D7-brane for O$1^-$, cannot support the end of a D-string.}
The variants we called $\hatplane{1}{-}$ and
$\hatplane{0}{-}$ therefore have the {\em same} charge (and
tension) as the original torsion-free planes. It would be
interesting to investigate whether (and how) the gauge theories on
D1-branes and D0-branes in the presence of these exotic
orientifold variants differ from the ordinary $SO(2n)$ theories on
O$1^-$ and O$0^-$.

The story becomes even more interesting for the lower O$p^+$
planes, as we will see next.

\subsection{Equivalent orientifolds}

O$p^+$ planes correspond to O$p^-$ planes with non-trivial $H$-torsion.
This will modify the AHSS for $p\leq 3$ through the differential $d_3$.
In particular, for the O$3^+$ plane
\be
\label{d3}
 d_3: H^0(\RP{5},\Z) \stackrel{\cup H}{\longrightarrow}
 H^3(\RP{5},\widetilde{\Z})
\ee
is a surjective map from $\Z$ to $\Z_2$. $G_3$ torsion therefore
lifts to a trivial class, and the graded complex is simply
$\mbox{Gr}KR^{-1}_\pm(\SSS^{6,0})=\Z$. As with the O$5^+$ plane,
the extensions are trivial, so the charge doesn't shift.
However, the new feature here is that, since the two
$G_3$ fluxes are equivalent in K-theory,
the two variants O$3^+$ and $\tildeplane{3}{+}$
are equivalent as well. Physically this can be understood as follows.
Equation (\ref{d3}) implies the existence of a constant 0-form
$\omega_0$, such that
\be
 G_3 \;=\; H\cup \omega_0  \;,
\ee
and the two O$3^+$ variants are related by a shift
$\omega_0 \rightarrow \omega_0 +1$. On the other hand, the NSNS
and RR 3-torsions of the two variants
are related by an $SL(2,\Z)$ transformation in Type
IIB string theory given by \cite{Witten_Baryon,HK}
\be
 T = \left[
 \begin{array}{cc}
 1 & 1 \\
 0 & 1
 \end{array}
 \right]\;,
\ee
which also shifts the axion $C_0\rightarrow C_0+1$. We therefore
identify $\omega_0$ with $C_0$.

Note that this transformation is global;
in a background with several O$3^+$ planes, it changes all of
them into $\tildeplane{3}{+}$ planes.
In particular, a configuration with two O$3^+$ planes at opposite points
on a circle is not equivalent to a configuration with one O$3^+$ and one
$\tildeplane{3}{+}$. This is also consistent with the fact that
the these configurations are related by T-duality to an O$4^+$ plane
and an $\tildeplane{4}{+}$ plane, respectively (fig.~3).
The latter are
inequivalent, since they differ by a ($G_2$) torsion class which
is not trivial in K-theory.

For $p\leq 2$ we similarly find that
$G_{6-p}$ torsion can be written as
\be
\label{trivial_torsion}
 G_{6-p} = H \cup \omega_{3-p} \;,
\ee
where $\omega_{3-p}\in H^{3-p}(\RP{8-p})=\Z_2$
(appropriately normal or twisted), so O$p^+$ and
$\tildeplane{p}{+}$ are equivalent in K-theory.
For $p=2$ there is again an interesting physical picture.
Consider the lift of O$2^+$ to M-theory \cite{Sethi,HK}. This corresponds
to an OM$2^+$ plane and and OM$2^-$ plane located at opposite points
on the circle.\footnote{Two OM$2^-$ planes correspond to an O$2^-$ plane,
and two OM$2^+$ planes correspond to an $\tildeplane{2}{-}$ plane}
The $\tildeplane{2}{+}$ plane is then simply
gotten by exchanging the positions of the two OM2 planes,
which is equivalent to the shift $x^{10}\rightarrow x^{10} \pm \pi$.
On the other hand,
the M-theory shift $x^{10}\rightarrow x^{10} + \xi$
($-\pi\leq\xi\leq\pi$) in flat space
reduces to a $U(1)$ gauge transformation in Type IIA string
theory,
\be
\label{two_plane_trans}
 \delta C_1 = d\xi \;, \; \delta C_3 = {B\over 2\pi} \wedge d\xi \;.
\ee
In the OM2 background we require
$\xi(-x^i)=-\xi(x^i)$, where $x^i$ ($i=3,\dots ,9$) are the coordinates
transverse to the O$2^+$-plane. In particular, at the location of the
orientifold plane $\xi(0)=0$ or $\pi$ (mod $2\pi$) only,
and this labels the discrete choice of the O$2^+$ variant.
Comparing (\ref{two_plane_trans}) with (\ref{trivial_torsion}) we are
led to identify $\omega_1$ with flat RR 1-forms
$C_1\in H^1(\RP{6},\widetilde{\Z})=\Z_2$.

In this case there are two possible non-trivial transformations for a pair
of O$2^+$ planes at opposite points on a circle,
\be
\xi_1(x^9)=\left\{
\begin{array}{rrr}
 -\pi & \;\;\mbox{for}\;\; & -\pi<x^9<0 \\
 \pi & \;\;\mbox{for}\;\; & 0<x^9<\pi
\end{array}\right.\;,
\qquad
\xi_2(x^9)=x^9\;.
\ee
In the first case both are transformed into $\tildeplane{2}{+}$ planes,
but in the second case only one (at $x^9=\pi$) is transformed.
The configuration with one O$2^+$ and one $\tildeplane{2}{+}$ is
therefore equivalent to the one with two O$2^+$ planes.
This is unlike the $O3^+$ case, since here the T-dual configurations
are $\tildeplane{3}{+}$ and O$3^+$, respectively (fig.~3),
which are equivalent.
\begin{figure}
\label{Tduality}
\centerline{\epsfxsize=3in\epsfbox{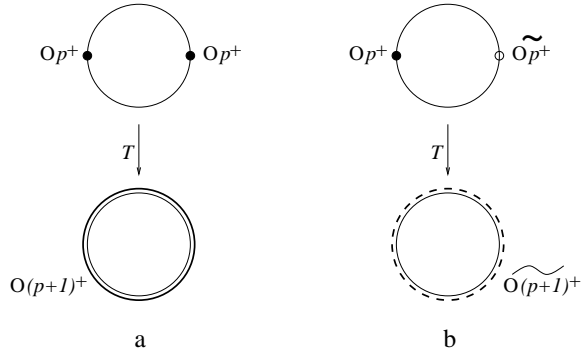}}
\medskip
\caption{T-duality relates (a) a configuration of two O$p^+$-planes
at opposite points on a circle to a single O$(p+1)^+$-plane
wrapping the circle, and (b) a configuration of one O$p^+$ and one
$\protect\tildeplane{p}{+}$ to a wrapped $\protect\tildeplane{(p+1)}{+}$.
The two configurations are equivalent for $p\leq 2$.}
\end{figure}

In general, the third differential $d_3$ will not correspond to just the
cup product with $H$, but the intuition remains.  If a field
strength $G_n$ can be written as a derivative $d_m\,C_{n-m}$,
then it can be trivialized by a gauge transformation, and does not
truly index a new RR field configuration.

\subsection{Unphysical orientifolds}

For O$p^+$ planes with $p\leq 2$ there is one more modification of the AHSS
due to the differential $d_3$. The map
\be
d_3:  H^{2-p}(\RP{8-p}) \stackrel{\cup H}{\longrightarrow} H^{5-p}(\RP{8-p})
 = \Z_2 \;,
\ee
where the cohomologies are appropriately normal or twisted (\ref{moreRR}),
implies that $G_{2-p}$ is not $d_3$-closed, but rather
\be
 d_3 G_{2-p} = \omega_{5-p}\in \Z_2\;.
\ee
If $\omega_{5-p}$ is the non-trivial element of $\Z_2$
then the corresponding class in $H^{2-p}$ cannot be lifted to
K-theory.
In particular, for the O$0^+$ and O$1^+$ planes $G_{2-p}$ torsion
is completely obstructed, which means that the variants
$\hatplane{0}{+}$ and $\hatplane{1}{+}$
(as well as the composites $\hattildeplane{0}{+}$
and $\hattildeplane{1}{+}$) do not exist.

For the O$2^+$ plane the above map is $\Z\rightarrow\Z_2$,
which implies that odd values of $G_0/2\pi$ are not allowed.
We will see in the next section that this can also be understood
in terms of an anomaly in the three-dimensional gauge theory on
D2-branes in an O$2^+$ background with odd $G_0/2\pi$.

It would be interesting to find a physical interpretation
for these obstructions, analogous to the interpretations we
gave for the equivalences before.

\section{The orientifold 6-plane}

\subsection{K-theory}

In the previous section we used the AHSS to obtain new insight
into how the cohomological classification of RR fluxes in
orientifold backgrounds is modified by K-theory. In particular, we
saw that independent RR fluxes may become correlated, certain RR
fluxes become trivial, and others are obstructed. For the
O6-planes the latter features are absent for dimensional reasons.
On the other hand we were not able to determine the correlation of
the two RR fluxes $G_2$ and $G_0$.

Luckily, representative gauge bundles for the relevant K-groups,
$KH(\SSS^{3,0})$ and $KR(\SSS^{3,0})$,
can easily be worked out.
Start with a flat Type IIA background, and consider complex vector
bundles over $\SSS^2$. These are classified by
$K(\SSS^2) = K(pt.) \oplus \widetilde{K}(\SSS^2) = \Z \oplus \Z$,
where the first term corresponds to the rank of the bundle, and
the second term to its topological ($ch_1$) class. 
Physically these correspond
to D8-branes ($G_0$ flux) and D6-branes ($G_2$ flux), respectively.
A convenient basis of generators to work with is $x=(1,0)$ and
$y=(1,1)$. The former corresponds to a constant rank 1 bundle $v_0$,
and the latter to the basic rank 1 (monopole) line bundle with
$ch_1 = 1$. This bundle can be constructed by assigning
trivial bundles $v_+$ and $v_-$ to the upper and lower
hemishpheres, respectively, with a transition function along the
equator,
\be
\label{transition}
v_+(\theta)=g(\theta) v_-(\theta) \;.
\ee
The
transition function $g(\theta)$ is a map $\SSS^1\rightarrow U(n)$,
which is classified by a winding number $w\in\pi_1(U(n))=\Z$. The
basic line bundle has $w=ch_1=1$.

Now consider the action of the involution on the two generators.
On the constant bundle the action is
\be
 v_0 \longrightarrow \bar{v_0}
\ee
if the involution squares to 1, and
\be
 v_0 \longrightarrow \J\bar{v_0}
\ee
if it squares to $-1$. Consequently $x$ is a generator for $KR(\SSS^{3,0})$,
whereas $2x$ is a generator for $KH(\SSS^{3,0})$.
The action on the line bundle is given by
\be
\begin{array}{rcc}
v_+(x) & \longrightarrow & \bar{v}_-(-x) \\
v_-(x) & \longrightarrow & \pm\bar{v}_+(-x) \;,
\end{array}
\ee
where the upper sign corresponds to the $KR$ projection and the lower sign
to the $KH$ projection.
On the equator this becomes
\be
\begin{array}{rcc}
v_+(\theta) & \longrightarrow & \bar{v}_-(\theta + \pi) \\
v_-(\theta) & \longrightarrow & \pm\bar{v}_+(\theta + \pi) \;.
\end{array}
\ee
Compatibility with (\ref{transition}) requires
\be
 g(\theta + \pi) = \pm\bar{g}(\theta)^{-1} = \pm g(\theta)^T \;,
\ee
which implies that $w\in 2\Z$ for $KR$, and $w\in 2\Z +1$ for $KH$.
Thus $y$ is a generator for $KH(\SSS^{3,0})$, whereas only
$2y$ is a generator for $KR(\SSS^{3,0})$.

The results are summarized as follows:
\be
\begin{array}{llll}
\mbox{Covering Space}: & K(S^2)=\Z\oplus \Z &
   \;\;\mbox{generators}\;\; x=(1,0) & \; y=(1,1) \\
 \mbox{O}6^-\mbox{plane}: & KH(S^{3,0})=\Z\oplus \Z &
   \;\;\mbox{generators}\;\;  2x=(2,0) & \; y=(1,1) \\
 \mbox{O}6^+\mbox{plane}: & KR(S^{3,0})=\Z\oplus\Z &
   \;\;\mbox{generators}\;\;  x=(1,0) & \; 2y=(2,2) \;.
\end{array}
\ee
A unit $G_2$ flux, {\em i.e.} a bulk D6-brane, in the reduced space
corresponds to the element $2y-2x=(0,2)$. We therefore see that
half-odd-integral $G_2$ flux, {\em i.e.} a fractional D6-brane,
is required for O$6^-$ when $G_0$ is odd, and forbidden when $G_0$ is even.
In the former case the gauge group on D6-branes is $SO(2n+1)$,
and we denote the orientifold plane $\tildeplane{6}{-}$.
For O$6^+$ $G_2$ flux is always integral, and independent of $G_0$.
The picture is similar to the one for the lower O$p$-planes,
except that instead of the charge quantum being correlated with
a RR torsion class (for O$p^-$), it is corellated with the parity of
an integral class.

\subsection{Anomalies}

The above conclusion can also be arrived at by considering
the world-volume theory on probe D-branes.
It was shown in \cite{Sugimoto} that a Chern-Simons (CS) term which includes
$G_0$ is crucial for cancellation of an anomaly with respect
to large gauge transformations in the world volume theory of
D2-branes in an $\tildeplane{6}{-}$ background.
We shall review the argument, and extend it to the other cases.

Consider first a flat Type IIA background. The CS terms on
a D$p$-brane ($p$ even) are given by \cite{GHT}
\be
S_{CS}^{flat} = \int_{p+1} C\wedge\tr e^{F/2\pi}
-\int_{p+1}\frac{G_0}{2\pi} \, \omega_{p+1}(A),
\label{CS}
\ee
where $\omega_{p+1}(A)$ is the Chern-Simons $(p+1)$-form defined by
\begin{eqnarray}
\label{CSform}
d \omega_{2n-1}(A)=2\pi\cdot\frac{1}{n!\,(2\pi)^n}\tr F^n
\end{eqnarray}
up to exact forms. Here `$\tr$' denotes the trace in the
fundamental representation of the gauge group.
As is well known, the integral of a CS form
is not gauge invariant, but rather shifts by a $2\pi$ multiple
of the winding number of the gauge transformation.
Namely, for an element $g\in U(n)$ with winding $w\in\pi_{p+1}(U(n))=\Z$,
the gauge transformation $A\rightarrow g^{-1}Ag+g^{-1}dg$ induces
\begin{eqnarray}
\int_{p+1}\omega_{p+1}(A)\rightarrow \int_{p+1}\omega_{p+1}(A)+2\pi w \;.
\end{eqnarray}
Invariance of the path integral therefore requires
$G_0/2\pi\in\Z$ \cite{GHT}.

In orientifold backgrounds the CS terms are given by
\be
S_{CS} =
{1\over 2}\int_{p+1} C\wedge\tr e^{F/2\pi}
-{1\over 2}\int_{p+1}\frac{G_0}{2\pi} \, \omega_{p+1}(A),
\label{CS2}
\ee
where now the traces are in the fundamental representation
of $G=O(n)$ or $USp(2n)$, and the factor of $1/2$
is needed to correctly reproduce the action (\ref{CS}) when the
D-branes move away from the orientifold plane.
The cases of interest are $p=2$ and $6$, as these are the only cases
(in ten dimensions) with integral winding, $\pi_3(G)=\pi_7(G)=\Z$.
The integrals of the corresponding CS forms defined in (\ref{CSform})
transform as follows
\begin{eqnarray}
\label{3variation}
 \delta \int\omega_3(A) & = & \left\{
 \begin{array}{ll}
 4\pi w & \;\;\mbox{for}\;\;O \\
 2\pi w & \;\;\mbox{for}\;\;USp
 \end{array}\right. \\
\label{7variation}
 \delta \int\omega_7(A) & = & \left\{
 \begin{array}{ll}
 2\pi w & \;\;\mbox{for}\;\;O \\
 4\pi w & \;\;\mbox{for}\;\;USp \;,
 \end{array}\right.
\end{eqnarray}
where $w$ is the appropriate winding number of the gauge transformation.
The relative factors of 2 can be understood by considering $USp$ and $O$
as projections of $U$. The generator of
$\pi_3(U)$ is associated with the instanton bundle on $\SSS^4$
({\em i.e.} the generator of $\widetilde{K}(\SSS^4)$).
Since this bundle is pseudoreal it survives the projection to $USp$,
but must be doubled to get the generator for $O$.
For $\pi_7$ the situation is reversed, since the associated bundle
on $\SSS^8$ is real.

There is an additional contribution to the CS action from the fermion
determinant $\det(iD\!\!\!\!/)$ known as the parity anomaly.
It was shown in \cite{R} that a complex fermion in a representaion
$R$ of the gauge group contributes to the three-dimensional CS term
as follows
\begin{eqnarray}
S^{\mbox{\tiny 1-loop}}_{CS}
=\half\frac{m}{|m|}\int \omega_{3}^{R}(A) \;,
\label{3dim}
\end{eqnarray}
where $m$ is the mass of the fermion. A similar computation in
seven-dimensional gauge theory gives 
\begin{eqnarray}
S^{\mbox{\tiny 1-loop}}_{CS}
=\half\frac{m}{|m|}\int \omega_{7}^{R}(A) \;.
\label{7dim}
\end{eqnarray}

Consider first $N$ D2-branes in the presence of an O6-plane and
$n$ D6-branes. The case of O$6^-$ was considered in \cite{Sugimoto}. In
this case the gauge group on the D2-branes is $USp(2N)$, and the
global symmetry corresponding to the D6-brane gauge group is
$O(2n)$ for O$6^-$ and $O(2n+1)$ for $\tildeplane{6}{-}$. The
spectrum of (complex) fermions in the D2-brane world-volume theory
consists of a state in the bi-fundamental representation
$(\funda\,,\funda\,)$, as well as a pair of states in the
symmetric representation of the gauge group $(\symm\,,{\bf 1})$,
and a pair in the antisymmetric representation $(\asymm\,,{\bf
1})$. Invariance (mod $2\pi$) of the sum of (\ref{CS2}) and
(\ref{3dim}) therefore requires $G_0/2\pi$ to be even for O$6^-$,
and odd for $\tildeplane{6}{-}$. For O$6^+$ the D2-brane gauge
group is $O(2N)$ (or $O(2N+1)$), and the global symmetry is
$USp(2n)$. The fermion representations are the same as above, so
in this case there is no parity anomaly. Furthermore, we see from
(\ref{3variation}) that $G_0/2\pi$ can be any integer. This is
exactly what we obtained in K-thoery.

The same result can also be obtained by considering
the world-volume theory on the D6-branes.
In this case there is a single complex fermion
in the adjoint representation of the gauge group.
The CS 7-forms defined in the adjoint representation
are related to those defined in the fundamental representaion by
\begin{eqnarray}
\int_{\Sigma_7}\omega_7^{\rm adj}&=&
(m-8)\int_{\Sigma_7}\omega_7 \quad\;\; \mbox{for} \;\; O(m) \\
\int_{\Sigma_7}\omega_7^{\rm adj}&=&
(2m+8)\int_{\Sigma_7}\omega_7 \quad \mbox{for} \;\; USp(2m)\;.
\end{eqnarray}
Together with (\ref{7variation}) we see that, in the first case,
invariance under a large gauge transformation
requires $G_0/2\pi\in 2\Z$ for O$6^-$ ($m$ even), and
$G_0/2\pi\in 2\Z + 1$ for $\tildeplane{6}{-}$ ($m$ odd).
There is no such correlation for O$6^+$, and
$G_0/2\pi$ is an arbitrary integer.

The same approach can be applied to the O$2$-plane with $G_0$ flux, for
which we only had partial K-theory results.
In particular, we determined that $G_0/2\pi$ had to be even
for O$2^+$, but we made no statement about O$2^-$.
The world-volume theory on D2-branes in the background of an
O2-plane contains an even number (four) of complex fermions in
the adjoint representation in all cases, so there is no
parity anomaly. Invariance under large gauge transformations
therefore requires $G_0/2\pi\in 2\Z$ for O$2^+$
(and $\tildeplane{2}{+}$), but allows
any integer value for O$2^-$ (and $\tildeplane{2}{-}$).
This implies, in particular, that the $G_6$ charge quantum is not
correlated with the value of $G_0$ (only with the torsion class
of $G_4$, as we saw in the previous section).

\subsection{Components of Type I vacuua}

In many ways the orientifold 6-plane was the motivation for this
investigation. Compactification of Type I string theory on
$\TT^3$, and T-duality, yields a background with eight O$6^-$
planes and sixteen D6-branes. On the other hand it is known that
the moduli space of Type I (or heterotic) vacua on $\TT^3$ has
additional disconnected components, with gauge groups of lower
rank \cite{Witten_novector,KRS,Keurentjes,Kac,FM}. 
One of our original goals was to describe these
different components in the T-dual picture, using D6-branes and
orientifold 6-planes.

Some of the additional components can indeed be described in this
way, by including both O$6^-$ and O$6^+$ planes, in much the same
way as one describes the eight-dimensional theory without vector
structure in terms of O$7^-$ and O$7^+$ planes
\cite{Witten_novector}. In the fully compact case the
seven-dimensional orientifold contains eight O6-planes, which are
located at the vertices of a cube. Since each plane can be either
O$6^-$ or O$6^+$, there are naively $2^8=256$ configurations.
However not all of them are consistent, and many of them are
related by the symmetry group of the 3-torus $SL(3,\Z)$.

Consider a configuration with $N_-$ O$6^-$ planes and $N_+$ O$6^+$
planes. Supersymmetry requires $N_-\geq N_+$, since otherwise
anti-D6-branes would be required for tadpole cancellation.
Furthermore, consistency of the orientifold projection requires
$N_-$ and $N_+$ to be even. For example, let us assume that the
plane at the origin of the cube is O$6^+$, and the seven others
are O$6^-$. If we denote the action of the projection on the CP
factors by $\{g,g_1,g_2,g_3\}$, where $g_i$ correspond to
translations along the torus, then we have
\be
\begin{array}{rcl}
 g^T &=& -g \\
 (gg_i)^T &=& +gg_i \;\;\;\;\; (i=1,2,3) \\
 (gg_ig_j)^T &=& +gg_ig_j \;\; (i\neq j) \\
 (gg_1g_2g_3)^T &=& +gg_1g_2g_3 \;.
\end{array}
\ee From the first three relations we deduce that \be
 \{g_i,g_j\}=0\;\;(i\neq j) \;.
\ee On the other hand this implies \be
\begin{array}{lll}
(gg_1g_2g_3)^T &=& -(gg_3g_2g_1)^T = -g_1^Tg_2^Tg_3^Tg^T \\
 & = & -(-gg_1g^{-1})(-gg_2g^{-1}g)(-g_3g^{-1})(-g) \\
 & = & -gg_1g_2g_3 \;,
\end{array}
\ee in contradiction with the last relation. The allowed values of
$(N_-,N_+)$ are thus $(8,0)$, $(6,2)$, and $(4,4)$ (fig.~4). In
the first case the configuration is unique and corresponds to the
original rank 16 component (fig.~4a). In the $(6,2)$ case the configuration
is unique up to $SL(3,\Z)$ transformations, and the rank is 8 (fig.~4b). 
In the $(4,4)$ case there is no gauge group, and there are actually two
$SL(3,\Z)$ equivalence classes. One corresponds to configurations
in which four O6-planes of the same type are co-planar (fig.~4c), and the
other to configurations where they are not co-planar 
(fig.~4d).\footnote{Since 
$M\in SL(n,\Z)$ satisfies $\mbox{det}M=1$, it follows that the
Levi-Civita tensor $\epsilon_{i_1\cdots i_n}$, and therefore also
the pseudo-scalar product of $n$ vectors $\epsilon_{i_1\cdots i_n}
v^{i_1}\cdots v^{i_n}$, is invariant under $SL(n,\Z)$. For $n=3$
this means that co-planarity is an $SL(3,\Z)$-invariant property.}
\begin{figure}
\centerline{\epsfxsize=5.5in\epsfbox{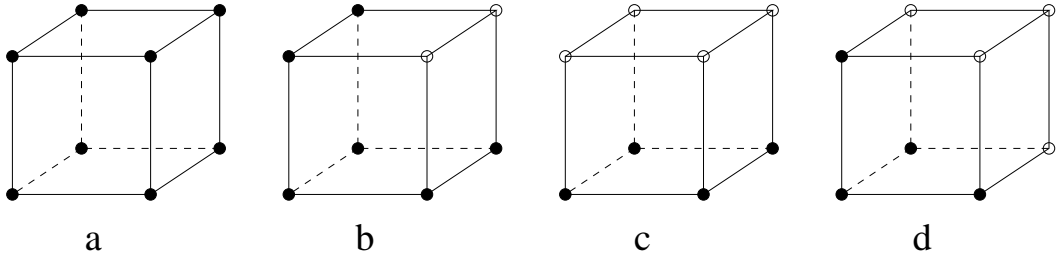}}
\medskip
\caption{Consistent $SL(3,\Z)$-inequivalent configurations of O6-planes:
(a) eight O$6^-$ planes (rank 16),
(b) six O$6^-$ planes and two O$6^+$ planes (rank 8),
(c)+(d) two inequivalent
configurations of four O$6^-$ planes and four O$6^+$ planes (rank 0).}
\end{figure}

Given the previous results, we now understand that the above
configurations require $G_0/2\pi$ to be even. In particular, they
exist for $G_0=0$. For $G_0/2\pi$ odd on the other hand, we must
replace O$6^-$ by $\tildeplane{6}{-}$, and change the number of
D6-branes appropriately to cancel the tadpole. This can only be
done supersymmetrically in the $(8,0)$ and $(6,2)$ configurations,
which yield new components with a rank 12 and a rank 5 gauge
group, respectively. 
(In the $(4,4)$ configurations the total charge
of the $\widetilde{\mbox{O}6}$-planes is $+2$, and can only be
cancelled by anti-D6-branes.) Of course since $G_0\neq 0$ in these
cases, there is a non-vanishing vacuum energy, and supersymmetry
is spontaneously broken.

\section{Conclusions and open questions}

In this paper we have demonstrated how K-theory generates very useful
information on the nature of discrete RR G-fluxes in string theory,
specifically for orientifold backgrounds. We saw that certain G-fluxes
are correlated, that some become identified due to discrete
transformations, and finally that some are disallowed outright.  A
very useful tool in this process was the Atiyah-Hirzebruch Spectral
Sequence and our generalization of it.  The AHSS ties together our
understanding of separate RR-fields (even torsion ones) in terms of
single K-classes.

    The AHSS was particularly powerful for our set of examples
because we already knew the relevant K-groups.  However, use of the
AHSS in general faces two obstacles.  The first obstacle involves
computing higher differentials in the sequence.  We were able to
circumvent this obstacle by using our exact knowledge of the
K-groups.  For complex K-theory, the $d_3$ differential has been
explicitly worked out \cite{DMW}.  For KR- and KH-theory, we did not
work these out, although (along the lines of \cite{Freed_Witten}) we
can see that adding the torsion H-fields to $d_3$ brings us from one to
the other. We would like to conclude by making some comments regarding
the other obstacle: the matter of extensions.

    One way to understand the extension problem is to realize that
while the graded complex generated by the AHSS yields the elements of
a K-group, it does not provide us with their additive structure.  We
propose that this additive structure can be extracted by thinking of
the K-group elements on the asymptotic sphere as limiting values of
generalized differentials in the bulk.  For our purposes, a general
differential form consists of a vector of differential forms of
different ranks with a differential action which can increase the rank
by more than one (for some ideas along these lines, we refer the
reader to the ongoing project in \cite{Freed,Freed_Hopkins}).

    The equation of motion for fields described by generalized
differentials involve the generalized derivative (naively a
combination of the $d_1,d_3$ and $d_5$ seen in the AHSS), which in 
general will
not be zero but can also give a K-class corresponding to the source of
the particular RR-fields.  The novelty here is that a K-class
corresponding to a $p$-form current $J_p$ 
could be a source for any field strength $G_m$ with
$m \leq p-1$, instead of just for $G_{p-1}$. Consistency
(commutativity) of the additive structure with the differential action
will put useful restrictions on the additive structure of the
generalized differential forms, which will in turn help determine the
group structure of the K-group at infinity.

    Another outstanding issue concerns the absolute charges of the
orientifold.  In this work we have only used K-theory to compute jumps
in the charges, and one can make use of the fact that K-theory
contributes and integral index to the path-integral to argue that
the jumps are anomaly free \cite{Moore_Witten}.  This leaves us with
one constant still to be determined, the bare charge of the
orientifold plane (for example, an $O3^-$-plane starts with charge
$-1/4$). 
Although some earlier work on this exists \cite{Mukhi}, there 
does not yet exist a satifactory answer within the K-theory framework.

\medskip

\noindent\underline{Note added}: The classification of consistent
and inequivalent O6-plane configurations (for $G_0=0$) in 
subsection 5.3 also
appeared in \cite{seven}.

\appendix

\section{AHSS in complex and real K-theory}

In this appendix we will summarize the notion of the
Atiyah-Hirzebruch spectral sequence (AHSS), which can be used to
compute K-theory groups which have a triangulation.  We will show
how this generalizes to real variants of K-theory.  The first
step in using the AHSS is to define an associated graded complex.

\subsection{Defining the Associated Graded Complex}

    Consider a space $X$, with a triangulation $\hat{X}$.  $X_p$
is the subset of this triangulation with simplices of maximum
dimension $p$.  For example the two-sphere, $\SSS^2$, has a
triangulation $\hat{X} = X_2$ equal to
\be
\{(x_2 x_3 x_4),(x_1 x_3 x_4),(x_1 x_2 x_4),(x_1 x_2 x_3)\}
\ee
and so
\be
\begin{array}{l}
X_1 = \{(x_1 x_2),(x_1 x_3), (x_1 x_4), (x_2 x_3), (x_2 x_4),
(x_3 x_4)\},\\ X_0 = \{x_1, x_2, x_3, x_4\}.
\end{array}
\ee
We are now going to use the triangulation to {\em filter} the
group K(X).  Define
\be
K_0(X) \doteq K(X),\; K_p(X)|_{1\,\leq\, p\, \leq Dim(X)} \doteq
\{x\,\epsilon\, K(X)\, |\; x\; {\rm trivial\ on}\; X_{p-1} \}.
\ee
Going back to our example $X=\SSS^2$, we have $K_0(\SSS^2) = \Z \oplus
\Z$. When we wrap a Dp-brane on this $S_2$, these two charges
correspond to Dp-brane charge and D(p$-2$)-brane charge.  Thus in
this example, $K_1(\SSS^2)=K_2(\SSS^2)= \Z$, reflecting the fact that
these groups capture only D(p$-2$)-brane charge.

    For a finite dimensional space X of dimension $N$, the $K_p$'s define
a {\em filtration},
\be
K = K_0 \,\supseteq\, K_1 \,\supseteq\, \ldots \,\supseteq\, K_N,
\ee
which makes $K(X)$ a {\em filtered complex}.  This allows us to
define an {\em associated graded complex}
\be
\mbox{Gr}K \doteq \bigoplus_{p=0}^{p=N} K_p/K_{p+1} \doteq
\bigoplus_{p=0}^{p=N} G_p.
\ee
Although $\mbox{Gr}K(X)$ doesn't capture all the structure of $K(X)$,
there is, as it turns out, a simple algorithm for computing $\mbox{Gr}K$
which usefully connects with much of the physicist's intuition
about fields as seen in cohomology.  Also note that we can easily
extend the notion of the associated graded complex $\mbox{Gr}K$ to any
other K-theory, such as $K^{-1}(X), KR(X), K_{\pm}(X)$, etc.

\subsection{Computing Gr$K$ using a spectral sequence}

    To compute $\mbox{Gr}K$, we are going to use a sequence of successive
approximations.  The first approximation to $G_p$ we can label as
$E_1^p$, and it consists of K-classes on the $p$-simplices
$\sigma^i_p$ trivial on their boundaries.  In other words,
\be
\label{e1}
E_1^p = K(\oplus_i \sigma_i^p, \oplus_i \dot{\sigma}_i^p) =
\oplus_i K(\sigma_i^p,\dot{\sigma}_i^p).
\ee
Of course, since $K(\sigma_i^p,\dot{\sigma}_i^p) = \widetilde{K}(\SSS^p)
= K^{-p}(pt)$ we get that
\be 
 E_1^p = C^p(X, K^{-p}(pt)).
\ee
Since we our using complex K-theory for now, only the
even-dimensional co-chains appear.

    Now the first approximation gives us a candidate K-classes on the
$\sigma_i^p$'s, but we need to make sure that they can be
extended to K-classes on $\hat{X}$ and thus $X$. The first step
is to extend a class on the $p$-skeleton to a class on the
($p+1$)-skeleton, i.e. take every element in $E_1^p$ and extend
it to the $\sigma_i^{p+1}$'s. Summing up the integers from each
component of the boundary $\dot{\sigma}_i^{p+1}$ yields, with
appropriate orientation, a winding number which if non-zero acts
as an obstruction to extending the co-chain to the
($p+1$)-skeleton. This is precisely the cohomology differential
which maps
\be
d_1: C^p(X, K^{-p}(pt)) \; \longrightarrow \; C^{p+1}(X, K^{-p}(pt)),
\ee
which implies that the next approximation to $G_p$ should be
\be
E_2^p = H^p(X, K^{-p}(pt)).
\ee

    To continue with this construction, we define elements
\be
\label{E2}
E_2^{p,q} = H^p(X, K^{q}(pt))\; {\rm with} \; E_2^p =
E_2^{p,-p}.
\ee
where $p$ ranges from 0 to the dimension of X, but $q$ is allowed
to range freely. With these elements, the whole approximation
scheme can then be refined recursively as follows.  We define
$d_r$ to be the obstruction to lifting a K-class in $E_r$ up $r$
steps. Formally,
\be
d_r: E_r^{p,q} \longrightarrow E_r^{p+r,q-(r-1)} \;.
\ee
We then define $E_{r+1}$ as the cohomology of the operator $d_r$
and repeat.  We have already seen what $E_1$, $d_1$ and $E_2$
correspond to.  In complex K-theory, only odd differentials
appear, and $d_3$ is the Steenrod square $Sq^3$.  For finite
dimensional spaces, this procedure terminates at some finite
order $n$.  We then have
\be
K_p(X)/K_{p+1}(X) = G_p(X) = E_n^{p,-p}(X).
\ee

    There are two features to this construction that are of particular
interest.  The first is that cohomology appears at the $E_2$
level in the scheme, hence we can view this construction of
K-classes as a sort of refinement of $H^{*}(X)$ with a different
addition structure.  The second is that the same coefficients
$E_r^{p,q}$ can naturally be used to compute the associated
graded complex for other K-theories (as long as there is no
involution on the space $X$) such as $K^{-s}(X)$ and $KO^{-s}(X)$
using a slight modification of the coefficients in eq (\ref{E2}).
For example
\be
G^{-s}_p(X) \doteq K^{-s}_p(X)/K^{-s}_{p+1}(X) =
E^{p,-(p+s)}_n(X).
\ee
Note that the differentials $d_r$ all map from the $E^{p,q}$'s
for $K^{-s}(X)$ to those for $K^{-s+1}(X)$ just as the regular
cohomology differential $d$ maps co-chains in $C^s(X)$ to
co-chains in $C^{s+1}(X)$.

\subsection{Extending the AHSS to K-theories with freely acting involutions}

    In order to extend the AHSS construction above to K-theories
with a freely acting spatial involution, e.g. $KR^{-s}(\SSS^{p,0})$, we
need to make some restrictions on the triangulation, namely that every
simplex $\sigma_i^p$ is mapped by the involution $\chi$ to another
simplex $\overline{\sigma_i}^p$ such that
\be
\label{exclude} \sigma_i^p\;\cap\;\overline{\sigma_i}^p = \emptyset\;.
\ee

As an example, we can give a simple such triangulation for the
$\SSS^{p,0}$.  Start with the set of points $\{x^1_{\pm},\ldots
,x^p_{\pm}\}$.  The simplices that make up the triangulation consist
of all sets $S$ of these points (with appropriate orientations)
such that
\be
x^i_{\pm} \,\in\, S \;\Rightarrow\; x^i_{\mp} \,\notin\, S.
\ee
For example, $(x^1_+\, x^2_-\, x^3_+)$ is a 2-simplex in this
triangulation, but $(x^1_+\, x^2_-\, x^1_-)$ is not.  We then define
the involution to to take
\be
\chi : \; (x^i_{\pm}, \ldots , x^j_{\pm}) \longrightarrow (x^i_{\mp},
\ldots , x^j_{\mp}).
\ee
which clearly satisfies eq.\ref{exclude}.

    Given a proper triangulation then, we can continue to follow the
procedure above, except now there should be a slight modification
of the $E_1$ term given in eq.\ref{e1}.  For example, for
computing $KR$-groups we use
\be
E_1^p = KR(\oplus_i (\sigma_i^p \oplus \bar{\sigma}_i^p), \oplus_i
(\dot{\sigma}_i^p \oplus \dot{\bar{\sigma}}_i^p)) = \oplus_i
KR(\sigma_i^p \oplus \bar{\sigma}_i^p,\dot{\sigma}_i^p \oplus
\dot{\bar{\sigma}}_i^p).
\ee
At this point it useful to bear in mind that generator of
$K(\sigma^i,\dot{\sigma}^i)$ is real (or pseudo real) for $i\,
=\, 0\, mod \, 4$ but not for $i\, =\, 2\, mod\, 4$ (the
conjugate spinors have opposite chirality) . This implies that for
$KR$- and $KH$-groups,
\begin{eqnarray}
\label{newe1}
E_1^{p,q}\; =\; C^p(X|_{\chi}, \Z) & {\rm for} & q\, =\, 0\, mod\, 4 \\
E_1^{p,q}\; =\; C^p(X|_{\chi}, \widetilde{\Z}) & {\rm for} & q\, =\, 2\,
mod\, 4 \\
E_1^{p,q}\; =\; 0 & {\rm for} & q\; {\rm odd}
\end{eqnarray}
where $\widetilde{Z}$ refers to the {\em local coefficients} on X
which switch signs with the map $\chi$, i.e., the co-chain takes
opposite value for opposite simplices in the covering space $X$.

   The differential $d_1$ is identical with the usual differential
operator, so $E_2$ gives twisted and un-twisted cohomologies. On
the other hand, we can see from eq. \ref{newe1} that $d_3$ takes
twisted chains to un-twisted chains and vice-versa and so is not
the Steenrod square like in the complex case.  It provides for a
difference between KR- and KH-groups.

\subsection{Illustrative Example: $KR(\SSS^{6,0})$ and
$KH(\SSS^{6,0})$}

    To illustrate the AHSS in the context of KR-groups and KH-groups,
we can examine as an example the groups $KR(\SSS^{6,0})$ and
$KH(\SSS^{6,0})$.  The $E_2$ terms are:
\begin{equation}
\begin{array}{lcccc}
E_2^{\,0,\,0\,}\;(\SSS^{6,0}) &=& H^0(\RP{5},\Z) &=&  \Z\\
E_2^{\,1,-1}(\SSS^{6,0}) &=& 0 && \\
E_2^{\,2,-2}(\SSS^{6,0}) &=& H^2(\RP{5},\widetilde{\Z}) &=& 0 \\
E_2^{\,3,-3}(\SSS^{6,0}) &=& 0 && \\
E_2^{\,4,-4}(\SSS^{6,0}) &=& H^4(\RP{5},\Z) &=& \Z_2 \\
E_2^{\,5,-5}(\SSS^{6,0}) &=& 0 && \\
\end{array}
\end{equation}
There are two possible places where the differential $d_3$ can
act non-trivially:
\begin{eqnarray}
d_3:\; E_2^{\,0,\,0\,}\;(\SSS^{6,0}) = \Z\;\, \rightarrow
E_2^{\,3,-2}(\SSS^{6,0}) = \Z_2 \label{zero}\\
d_3:\; E_2^{\,1,-2}(\SSS^{6,0}) = \Z_2 \rightarrow
E_2^{\,4,-4}(\SSS^{6,0}) = \Z_2 \label{one}
\end{eqnarray}

    Let us look at the case (\ref{zero}) more closely.  The generator
of $E_2^{\,0,\,0}(\SSS^{6,0})$ is a bundle of rank one over the
0-skeleton. Being in $E_2$ (and thus $E_3$) means that it extends
to a rank one bundle over the 2-skeleton.  Let's try to extend it
to a rank one bundle over the 3-skeleton. Consider the 3-chain in
$C_3(\RP{5},\widetilde{\Z})$ which looks like an $\RP{3}$ inside
$\RP{5}$.  It has as a boundary a 2-chain made up of two
$\RP{2}$'s (which explains why $H_3(\RP{5},\widetilde{\Z}) = 0$). In
KR-theory, the constant rank one bundle over the $\RP{2}$'s
can be extended to the ``$\RP{3}$'', but in KH-theory the rank one
bundle over $\RP{2}$ has non-trivial first Chern class and so
cannot be extended to the ``$\RP{3}$''.  Looking at eq.
(\ref{zero}), this fact implies that $d_3$ takes $z \rightarrow
0$ in $KR(\SSS^{6,0})$ but that $z \rightarrow z \,mod\, 2$ in
$KH(\SSS^{6,0}).$  Further analysis of the same kind reveals that
$d_3$ is only non-trivial in eq. (\ref{one}) for $KH(\SSS^{6,0}).$
The final result is then:
\be
\begin{array}{rcl}
KR(\SSS^{6,0}) &=& \Z \oplus \Z_2 \\
KH(\SSS^{6,0}) &=& 2\Z.
\end{array}
\ee
where the factor of 2 denotes the fact that only even rank
constant bundles are allowed.

\vspace{0.5cm} \noindent {\bf Acknowledgments}

\noindent O.B. and E.G would like to thank M. Atiyah, A. Hanany, B. Kol,
and especially E. Witten for useful conversations.
S.S. would like to thank Y. Hyakutake and Y. Imamura for
useful discussions.
O.B. was supported in part by the DOE under grant no.
DE-FG03-92-ER40701, and by a Sherman Fairchild Prize Fellowship.
The work of E.G. was supported by the NSF under grant
no. 0070928 and Frank and Peggy Taplin.
The work of S.S. was supported in part by the Japan Society
for the Promotion of Science.

\end{document}